\documentclass[3p,times,procedia]{elsarticle}
\usepackage{nupha_ecrc}

\volume{00}

\firstpage{1}

\journalname{Nuclear Physics A}

\runauth{D. McGlinchey}

\jid{nupha}

\jnltitlelogo{Nuclear Physics A}

\usepackage{amssymb}
\usepackage{lineno}

\usepackage[figuresright]{rotating}

\usepackage[export]{adjustbox}% http://ctan.org/pkg/adjustbox
\usepackage{xspace}

\newcommand{\pp}{\mbox{$p+p$}\xspace}
\newcommand{\pau}{\mbox{$p$$+$Au}\xspace}
\newcommand{\pal}{\mbox{$p$$+$Al}\xspace}

\newcommand{\dau}{\mbox{$d$$+$Au}\xspace}
\newcommand{\heau}{\mbox{$^3$He$+$Au}\xspace}
\newcommand{\pdheau}{\mbox{$p/d/^3$He$+$Au}\xspace}
\newcommand{\cucu}{\mbox{Cu$+$Cu}\xspace}
\newcommand{\cuau}{\mbox{Cu$+$Au}\xspace}
\newcommand{\auau}{\mbox{Au$+$Au}\xspace}
\newcommand{\pbpb}{\mbox{Pb$+$Pb}\xspace}
\newcommand{\sqsn}{\mbox{$\sqrt{s_{_{NN}}}$}\xspace}
\newcommand{\sqsntwo}{\mbox{$\sqrt{s_{_{NN}}}=200$~GeV}\xspace}
\newcommand{\pt}{\mbox{$p_T$}\xspace}

\newcommand{\pio}{\mbox{$\pi^0$}\xspace}

\newcommand{\vtt}{\mbox{$v_2\{2\}$}\xspace}
\newcommand{\vtte}{\mbox{$v_2\{2, |\Delta\eta|>2\}$}\xspace}
\newcommand{\vtf}{\mbox{$v_2\{4\}$}\xspace}

\begin{document}

\begin{frontmatter}



%% Instructions from Editor: Please use the following \dochead only in the preprint version (e-print arXiv etc.); 
%% use empty \dochead{} when submitting to Nuclear Physics A!
\dochead{XXVIth International Conference on Ultrarelativistic Nucleus-Nucleus Collisions\\ (Quark Matter 2017)}
%% e.g. \dochead{17th International Conference on Dynamical Processes in Excited States of Solids}

\title{PHENIX Overview}


\author{D. McGlinchey (for the PHENIX Collaboration)}

\address{University of Colorado Boulder}

\begin{abstract}

These proceedings present highlighted results from PHENIX shown at the Quark Matter 2017 conference.
%  on collectivity in small systems, electromagnetic probes, high-$p_T$ hadrons, and heavy flavor production

\end{abstract}


\end{frontmatter}

% \linenumbers

\section{Introduction}
\label{sec:intro}

For 16 years, the PHENIX experiment at the Relativistic Heavy Ion Collider (RHIC) has been furthering our understanding of the strongly coupled Quark Gluon Plasma (QGP) formed in heavy ion collisions over a broad range of topics. Observations of flow-like signatures in small collision systems, once thought to be necessary to understanding our baseline for ion-ion collisions, at both the Large Hadron Collider (LHC)~\cite{CMS:2012qk,Abelev:2012ola,Aad:2012gla} and RHIC~\cite{Adare:2013piz,Adare:2014keg,Adare:2015ctn,Aidala:2016vgl} have set off a vigorous new investigations into the conditions required to form a QGP. Over the last few years, RHIC has provided the highest luminosity samples of \auau collisions at \sqsntwo to date, as well as a suite of small collision systems with both a geometry scan of \pdheau collisions at \sqsntwo and an energy scan of \dau collisions. We are now beginning to see the fruits of these labors and their effects on our understanding of the QGP.

Emphasizing the versatility of RHIC, PHENIX presented new results spanning 8 collision systems and 5 collision energies at this conference. These results were presented in 11 parallel talks and 22 posters. These proceedings highlight a selection of new results from PHENIX in collective dynamics (Sec.~\ref{sec:flow}), EM probes (Sec.~\ref{sec:emprobes}), high-\pt hadrons (Sec.~\ref{sec:jets}), and heavy flavor (Sec.~\ref{sec:hf}) presented at Quark Matter 2017. 

%%%%%%%%%%%%%%%%%%%%%%%%%%%%%%%%%%%%%%%%%%%%%%%%%%%%%%%%%%%%%%%%%%%%%%%
%%%%%%%%%%%%%%%%%%%%%%%%%%%%%%%%%%%%%%%%%%%%%%%%%%%%%%%%%%%%%%%%%%%%%%%
%%%%%%%%%%%%%%%%%%%%%%%%%%%%%%%%%%%%%%%%%%%%%%%%%%%%%%%%%%%%%%%%%%%%%%%
\section{Collective Dynamics in Small Collision Systems}
\label{sec:flow}

\begin{figure}[htb]
	\centering
	\includegraphics[width=0.45\textwidth]{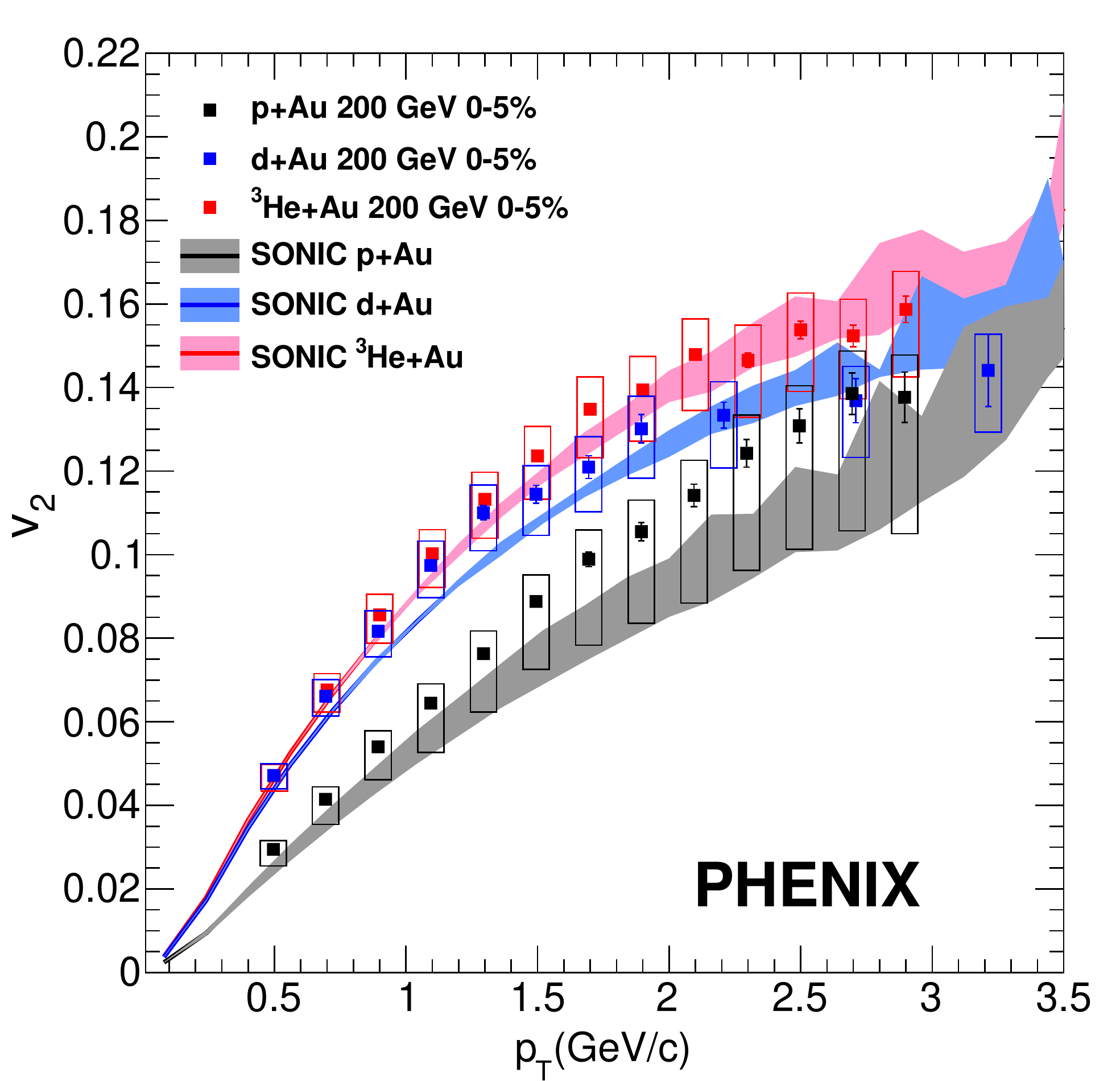}
	\includegraphics[width=0.45\textwidth]{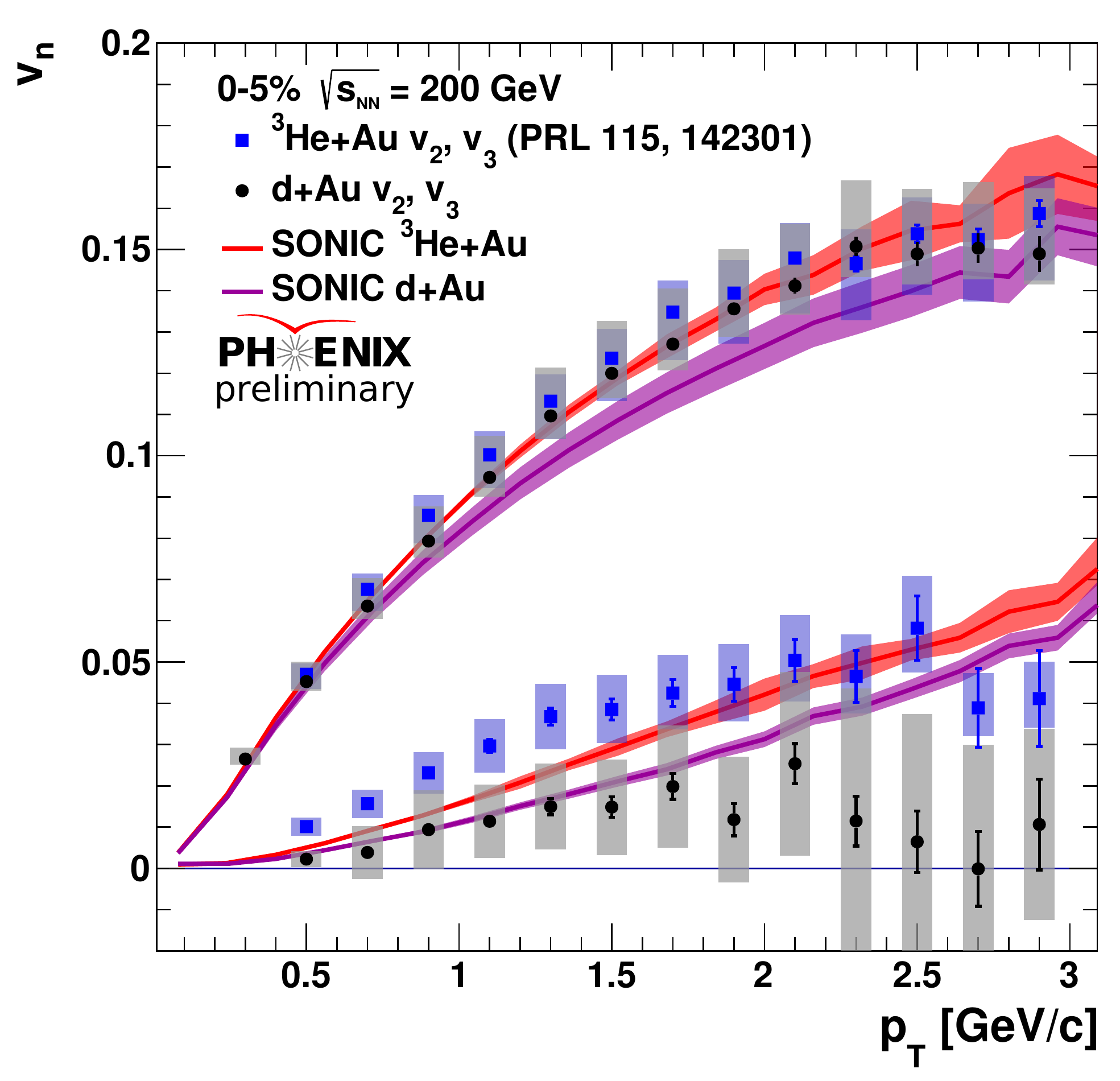}
	\caption{\label{fig:v2v3smallsys} (Left) The $v_2$ vs \pt in \pdheau collisions at \sqsntwo~\cite{Aidala:2016vgl} compared to calculations from SONIC. (Right) The $v_2$ and $v_3$ in $d/^3$He$+$Au collisions at \sqsntwo compared to calculations from SONIC.}
\end{figure}

PHENIX has furthered its investigation into the effects of initial geometry on flow signatures with the publication of $v_2$ vs \pt in \pau collisions at \sqsntwo~\cite{Aidala:2016vgl} and new preliminary measurements of $v_2$ vs \pt in \pal at the same collision energy. Combining these results with the previous measurements in \dau~\cite{Adare:2014keg} and \heau~\cite{Adare:2015ctn} provides stringent constraints on interpretations of the measured $v_2$ signal in small collision systems. If the final state $v_2$ arises from initial spatial correlations propagated to final state momentum correlations via hydrodynamics, a clear ordering between the flow strengths in three collision systems is expected, as laid out in Ref.~\cite{Nagle:2013lja}. Figure~\ref{fig:v2v3smallsys} shows the PHENIX measurements of $v_2$ in \pdheau at \sqsntwo~\cite{Aidala:2016vgl}, as well as the $v_3$ in \heau at \sqsntwo~\cite{Adare:2015ctn} and the preliminary result on $v_3$ in \dau at \sqsntwo from data taken in 2016. The $v_n$'s show a clear ordering between systems, with $v_2^{^3\mathrm{He}+\mathrm{Au}}\sim v_2^{d+\mathrm{Au}} > v_2^{p+\mathrm{Au}}\sim v_2^{p+\mathrm{Al}}$ (\pal not shown) and $v_3^{^3\mathrm{He}+\mathrm{Au}} > v_3^{d+\mathrm{Au}}$. This ordering is consistent with expectations from initial geometry + final state interactions. Further, calculations from SONIC~\cite{Habich:2014jna}, a hydrodynamical model, are in good agreement with the measurements.

\begin{figure}[htb]
	\centering
	\includegraphics[width=0.98\textwidth]{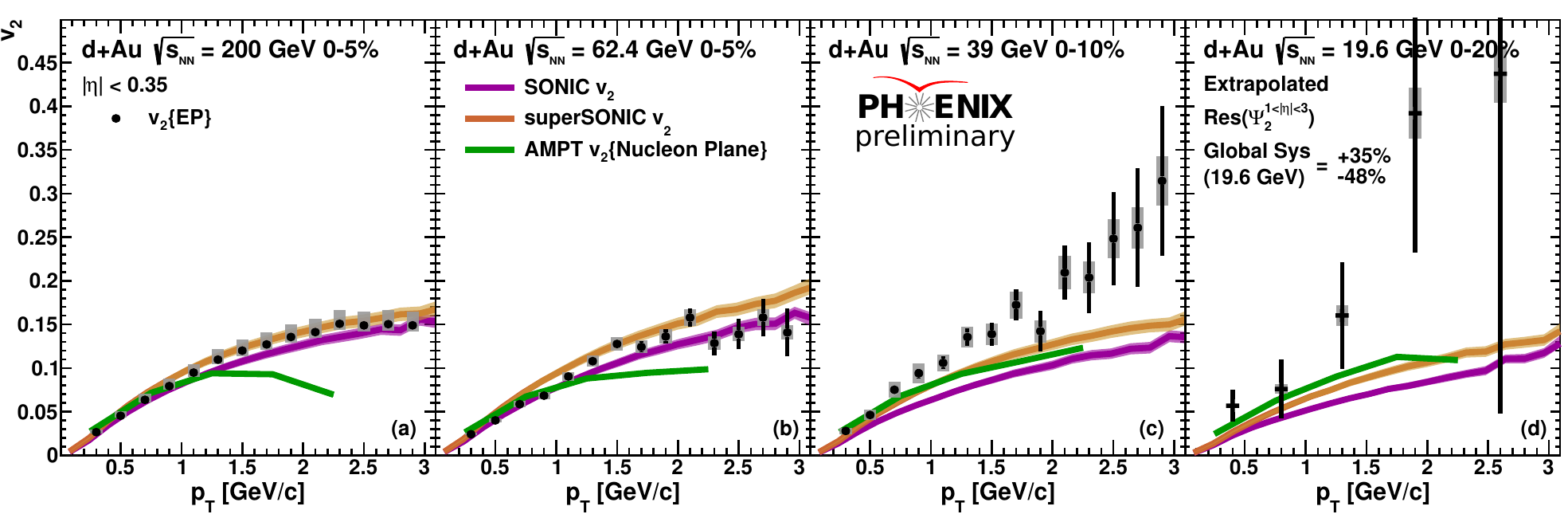}
	\caption{\label{fig:v2dautheory} The $v_2$ vs \pt in \dau collisions at \sqsn~=~200 (a), 62.4 (b), 39 (c), and 19.6 (d) GeV compared to calculations from SONIC, superSONIC, and AMPT in Ref.~\cite{Koop:2015trj}.}
\end{figure}

\begin{figure}[htb]
	\centering
	\includegraphics[width=0.98\textwidth]{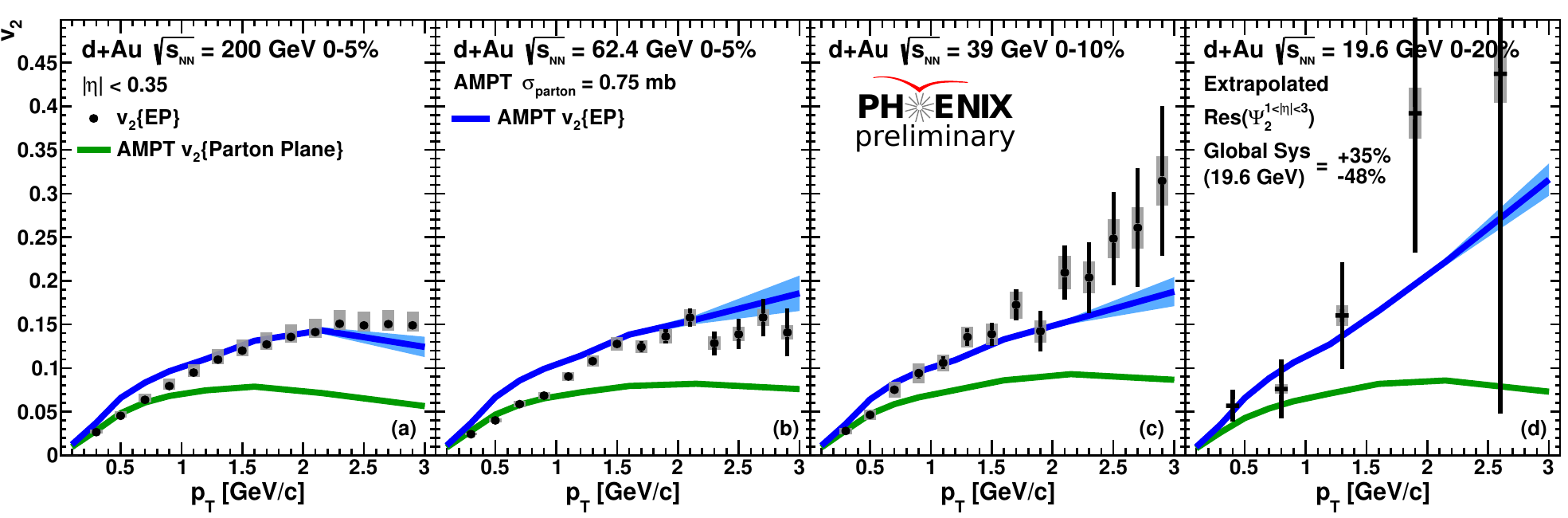}\\
	\includegraphics[width=0.8\textwidth]{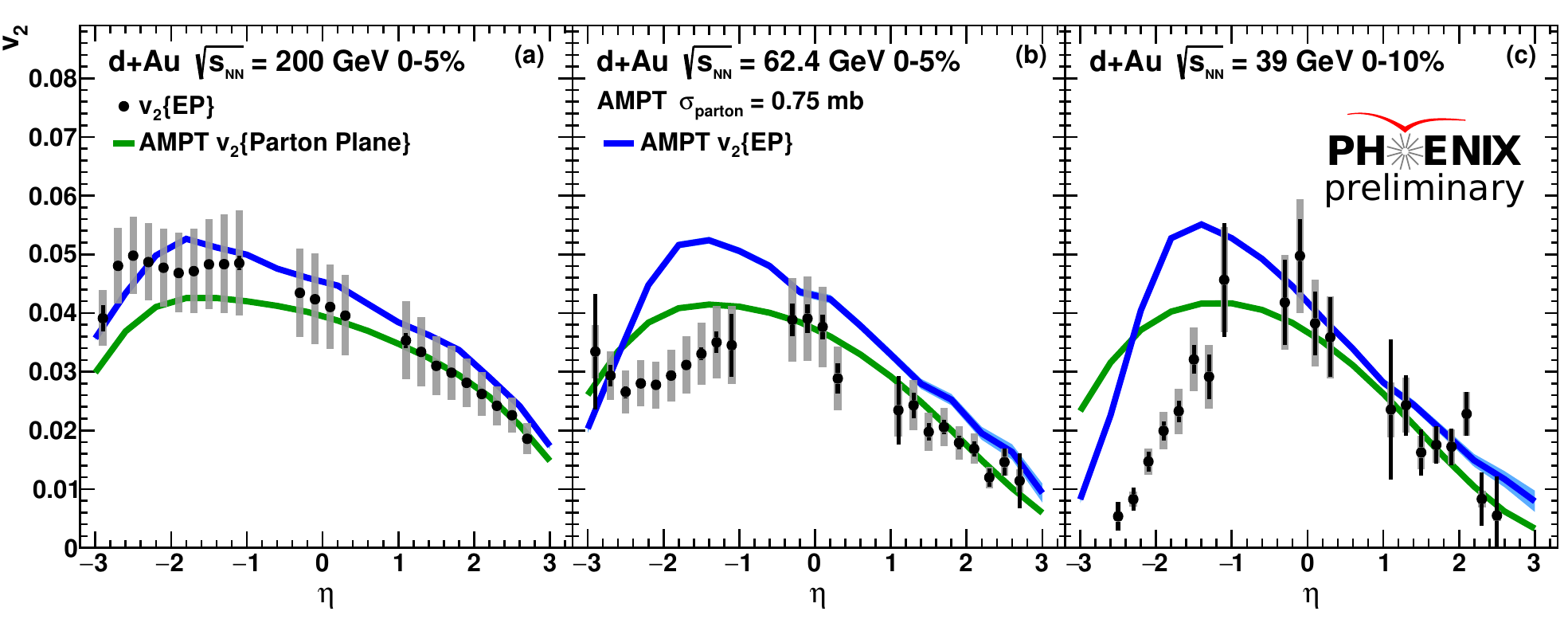}
	\caption{\label{fig:v2ptetadau} The $v_2$ vs \pt (Top) and $\eta$ (Bottom) in \dau collisions at \sqsn~=~200 (a), 62.4 (b), 39 (c), and 19.6 (d) GeV compared to calculations from AMPT.}
\end{figure}

In 2016 RHIC delivered a beam energy scan of \dau collisions at four different collision energies, \sqsn~=~200, 62.4, 39, and 19.6 GeV in order to investigate the onset of collectivity. From these data, PHENIX has made preliminary measurements of $v_2$ vs \pt and $\eta$ at all four energies, shown in Figs.~\ref{fig:v2dautheory} and~\ref{fig:v2ptetadau}. A clear $v_2$ signal which rises with \pt is observed at all four collision energies. The $\eta$ dependence shows a decrease of the $v_2$ at forward rapidity ($d$-going). At 200 GeV the backward rapidity (Au-going) is similar to the midrapidity result, however as the collision energy is decreased the $v_2$ at backward rapidity appears to collapse.

Figure~\ref{fig:v2dautheory} shows comparisons to hydrodynamic calculations from SONIC and superSONIC~\cite{Romatschke:2015gxa} as well as parton scattering from A Multiphase Transport Model (AMPT)~\cite{Lin:2004en}. Reasonable agreement between the data and the models is seen at low-\pt for all four energies. However, at 39 and 19.6 GeV, all three models underpredict the data for $\pt>1$ GeV/$c$. This is likely due to an increase in the non-flow contribution, which is not subtracted from the data and is not included in the calculations. The non-flow contribution is expected to increase with increasing \pt and with decreasing collision energy.

While the SONIC and superSONIC calculations do not include non-flow effects, AMPT, being a full event generator, in principle can include non-flow effects. Figure~\ref{fig:v2ptetadau} compares $v_2$ vs \pt and $\eta$ between data and two different calculations from AMPT. The green curves show $v_2$ calculated in AMPT relative to the parton participant plane ($v_2\{PP\}$). This result includes only the $v_2$ relative to the initial geometry, or what we refer to as flow. The blue curve shows $v_2$ relative to the event plane ($v_2\{EP\}$), mimicking the same method used in the experiment. This includes both flow effects, correlated to the initial geometry, and non-flow, which is not correlated to the initial geometry. The $v_2\{EP\}$ calculation shows much better agreement with the $v_2$ vs \pt measurement compared to $v_2\{PP\}$, particularly at higher \pt. The difference between the $v_2\{EP\}$ and parton plane results indicates that, at least in AMPT, non-flow plays a significant role at high-\pt and increases with decreasing collision energy. It is also interesting that the $v_2\{PP\}$ vs $\eta$ from AMPT shows a drop in the $v_2$ at backward rapidity at 62.4 and 39 GeV. This likely indicates some anti-correlation effects when measuring $v_2$ in an $\eta$ region near where the event plane is measured ($-3.9<\eta<-3.1$ in both AMPT and data).

\begin{figure}[htb]
	\centering
	\includegraphics[width=0.98\textwidth]{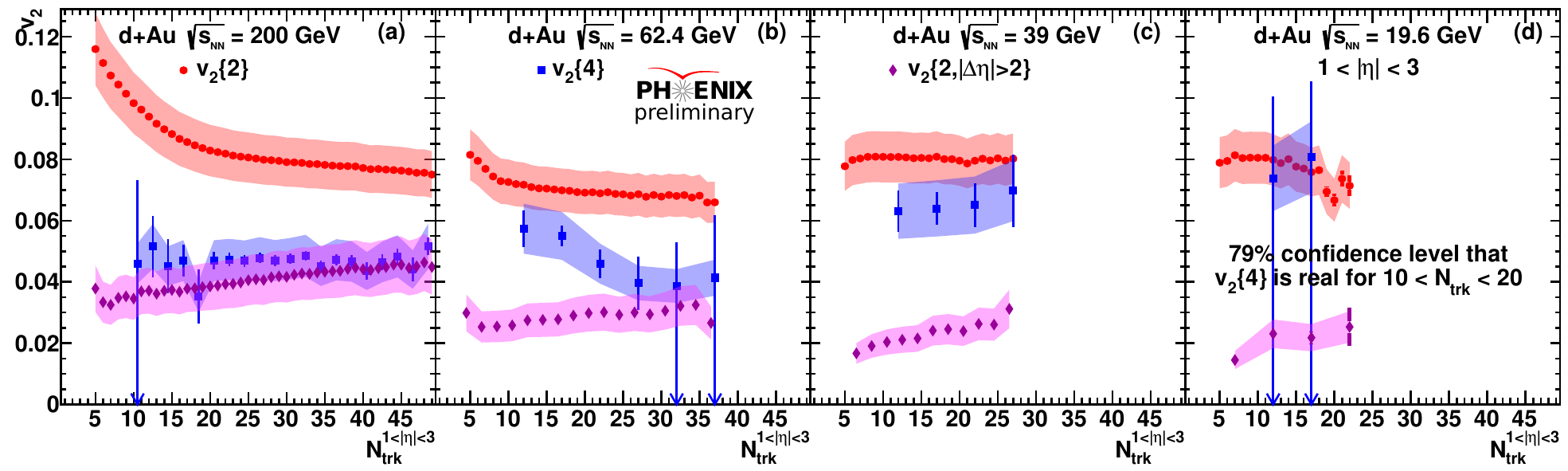}
	\caption{\label{fig:v24dau} The two and four particle cumulants as a function of number of tracks in \dau collisions at \sqsn~=~200 (a), 62.4 (b), 39 (c), and 19.6 (d) GeV.}
\end{figure}

For the first time in PHENIX, we have also made preliminary measurements of two and four particle cumulants in the \dau beam energy scan. We find a real \vtf, shown in Fig.~\ref{fig:v24dau}, at all four energies. This is a strong indicator of true multiparticle correlations in \dau collisions even at 19.6 GeV. We have also measured the two particle cumulant, \vtt, also shown in Fig.~\ref{fig:v24dau}. We find that the $\vtt > \vtf$, as expected from Gaussian fluctuations in the event-by-event $v_2$~\cite{Ollitrault:2009ie}. Since the \vtt is susceptible to non-flow contributions we apply a $|\Delta\eta|>2$ cut on track pairs, yielding \vtte, also shown in Fig.~\ref{fig:v24dau}. We find that $\vtte < \vtf$ and that \vtte decreases with decreasing collision energy, contrary to expectations. This can be understood in the following way. In this measurement, requiring the $\Delta\eta$ gap by construction correlates one particle at negative rapidity with one particle at positive rapidity. This means one has the product of the true $v_2$ at forward rapidity ($v_2^F$) and backward rapidity ($v_2^B$). In contrast, \vtt and \vtf, are a weighted average of $v_2^F$ and $v_2^B$. Since the $dN_{ch}/d\eta$ distribution is largest at backward rapidity, and the true $v_2$ is falling at forward rapidity, the \vtt and \vtf are more weighted towards the larger $v_2^B$ compared to the direct product in \vtte.

%%%%%%%%%%%%%%%%%%%%%%%%%%%%%%%%%%%%%%%%%%%%%%%%%%%%%%%%%%%%%%%%%%%%%%%
%%%%%%%%%%%%%%%%%%%%%%%%%%%%%%%%%%%%%%%%%%%%%%%%%%%%%%%%%%%%%%%%%%%%%%%
%%%%%%%%%%%%%%%%%%%%%%%%%%%%%%%%%%%%%%%%%%%%%%%%%%%%%%%%%%%%%%%%%%%%%%%
\section{Electromagnetic Probes}
\label{sec:emprobes}

\begin{figure}[htb]
	\centering
	\includegraphics[width=0.7\textwidth]{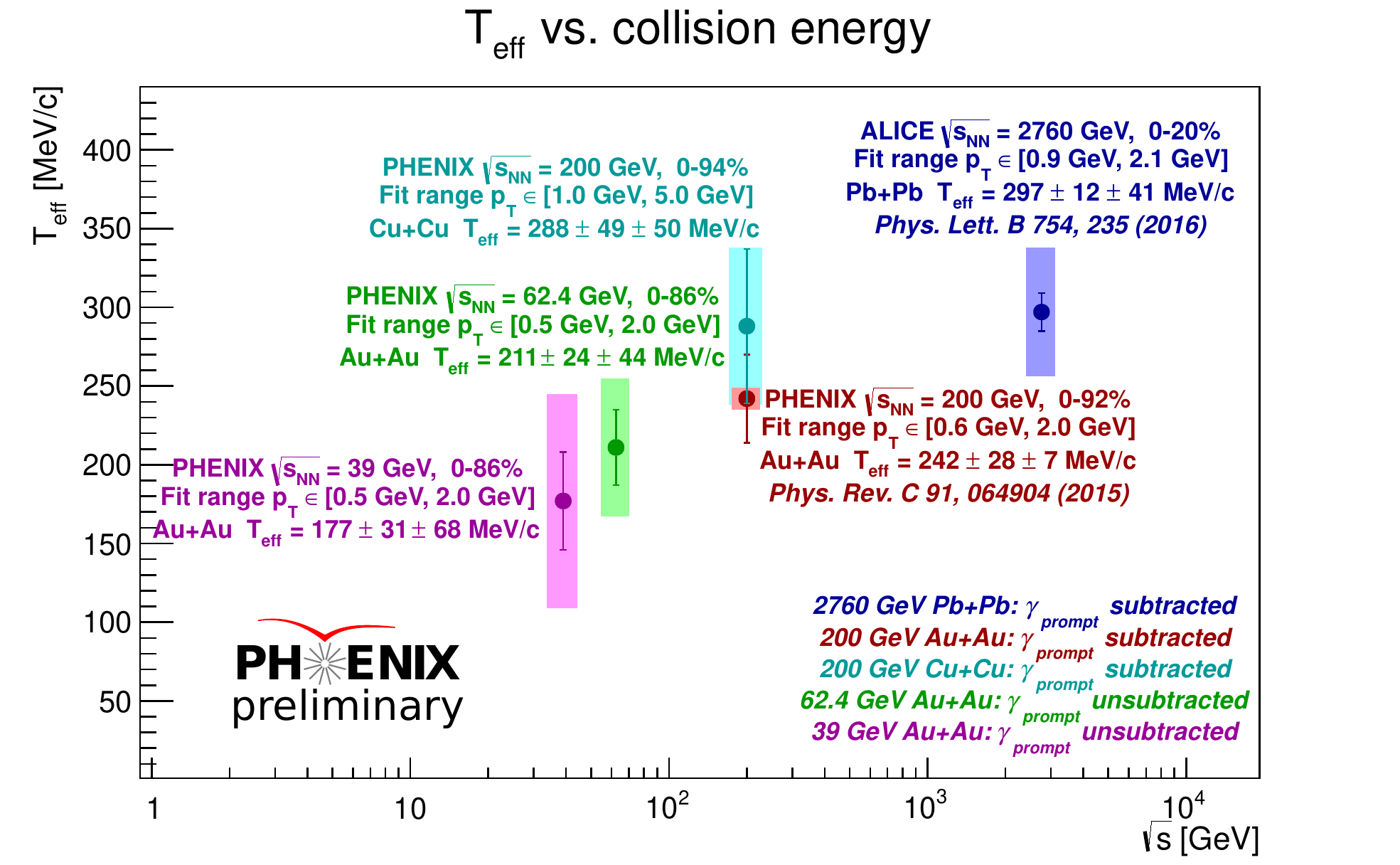}
	\caption{\label{fig:teff} The direct photon slope, $T_{eff}$ as a function of collision energy.}
\end{figure}

In order to gain further insight into the temperature and evolution of the QGP, PHENIX has made new preliminary measurements of direct photon production in \cucu collisions at \sqsntwo and \auau collisions at \sqsn~=~62.4 and 39 GeV. An excess of direct photon production is found in both systems. The inverse slope of the direct photon excess, $T_{eff}$, is shown in Fig.~\ref{fig:teff} alongside previous PHENIX results in \auau collisions at \sqsntwo and ALICE results in \pbpb at \sqsn~=~2.76 TeV. While the systematics on the preliminary results are large, they indicate a trend of increasing $T_{eff}$ with increasing collision energy. A reduction in systematic uncertainties is expected in the final results, which will provide more constraints on the energy dependence of $T_{eff}$.

%%%%%%%%%%%%%%%%%%%%%%%%%%%%%%%%%%%%%%%%%%%%%%%%%%%%%%%%%%%%%%%%%%%%%%%
%%%%%%%%%%%%%%%%%%%%%%%%%%%%%%%%%%%%%%%%%%%%%%%%%%%%%%%%%%%%%%%%%%%%%%%
%%%%%%%%%%%%%%%%%%%%%%%%%%%%%%%%%%%%%%%%%%%%%%%%%%%%%%%%%%%%%%%%%%%%%%%
\section{High-\pt Hadrons}
\label{sec:jets}

\begin{figure}[htb]
	\centering
	\includegraphics[trim={0 0 2cm 1.5cm},clip,width=0.32\textwidth]{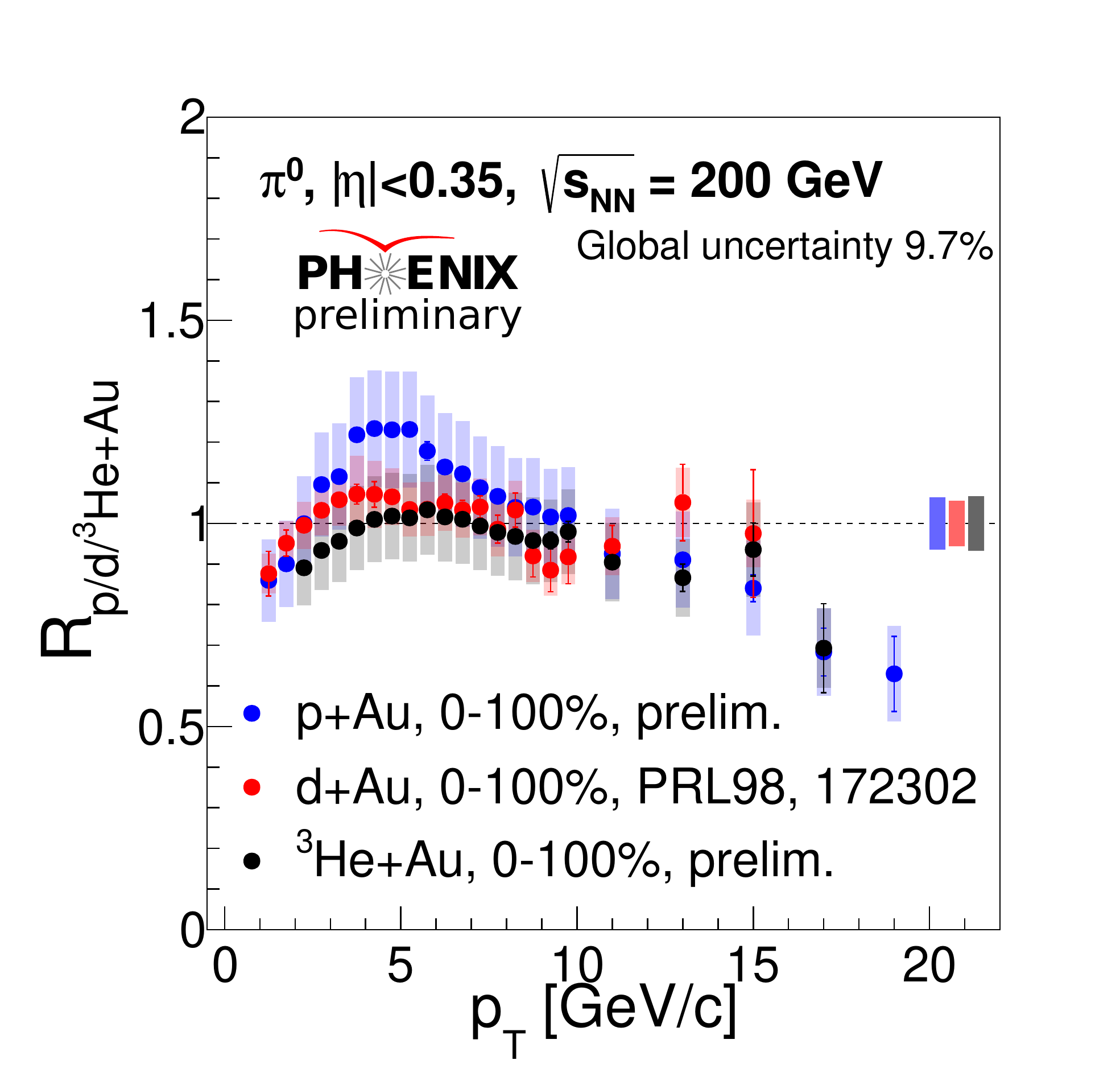}
	\includegraphics[trim={0 0 2cm 1.5cm},clip,width=0.32\textwidth]{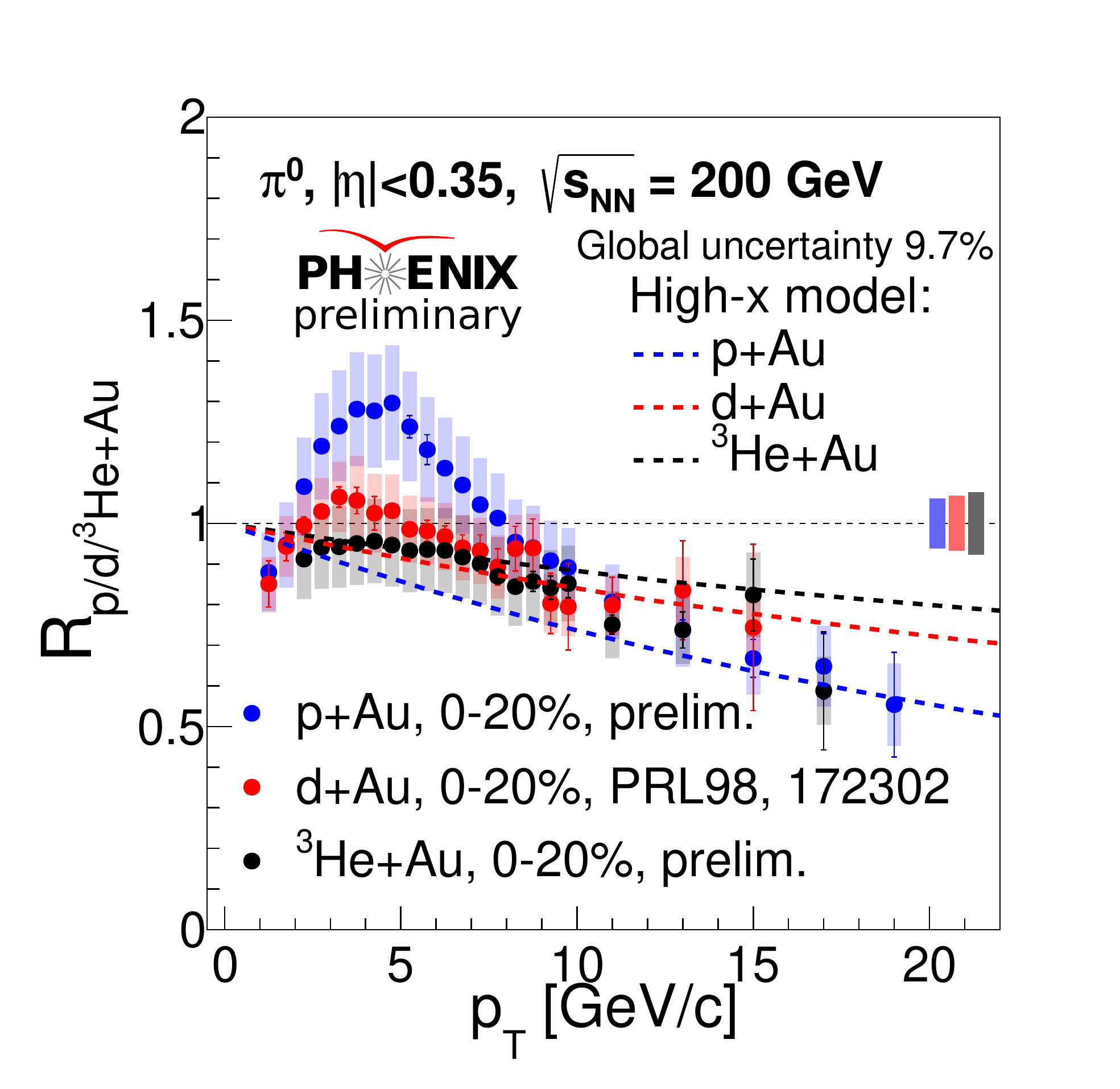}
	\includegraphics[trim={0 0 2cm 1.5cm},clip,width=0.32\textwidth]{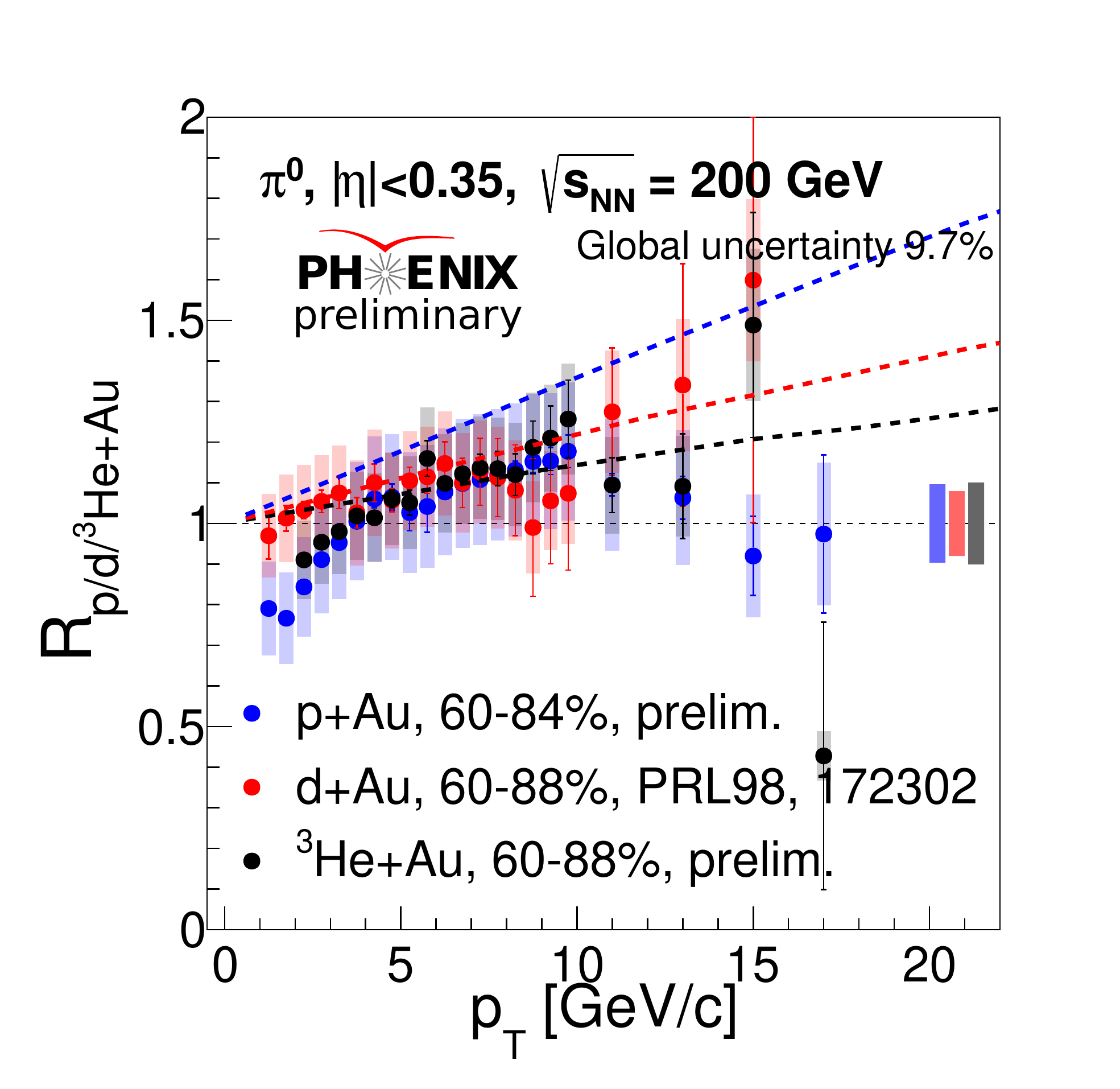}
	\caption{\label{fig:pi0smallsys} The $R_{AA}$ for $\pi^0$'s in MB (Left), central (Middle), and peripheral (Right) \pdheau at \sqsntwo. The dashed lines are calculations from Ref.~\cite{McGlinchey:2016ssj}.}
\end{figure}

New preliminary results on the modification of \pio production in \pau collisions completes a beam species scan of \pio production in small collision systems at \sqsntwo. The resulting nuclear modification factors are shown in Fig.~\ref{fig:pi0smallsys} as a function of \pt for minimum bias (MB), central, and peripheral \pdheau collisions at \sqsntwo. At $\pt\sim5$ GeV/$c$ a clear ordering is seen, with a peaked enhancement observed in \pau, a smaller enhancement seen in \dau, and little enhancement seen in \heau. This ordering may be due to different multiple scattering effects in the three collision systems. At high-\pt, a suppression is seen in central events while an enhancement is observed in peripheral events. This is reminiscent of the results on jet suppression in \dau collisions~\cite{Adare:2015gla}. It has been proposed that this effect could be due to fluctuations of the proton size, and that comparing results in \pdheau collisions would be an ideal test of that hypothesis~\cite{McGlinchey:2016ssj}. The model outlined in Ref.~\cite{McGlinchey:2016ssj} predicts that the \pau should have the largest enhancement (suppression) in peripheral (central) events, followed by \dau then \heau. The calculation is shown as dashed lines in Fig.~\ref{fig:pi0smallsys} for central and peripheral events. The calculation is in reasonable agreement with the data in central events for $\pt>10$ GeV/$c$, however no clear ordering is seen within the experimental uncertainties. In peripheral events, the calculation is in reasonable agreement with the \dau and \heau data, however the calculation expects the largest enhancement in \pau, which is not seen in the data.
% However, no clear ordering in central events is observed within the experimental uncertainties, and in peripheral events the \pau is consistent with no enhancement at high-\pt.

%%%%%%%%%%%%%%%%%%%%%%%%%%%%%%%%%%%%%%%%%%%%%%%%%%%%%%%%%%%%%%%%%%%%%%%
%%%%%%%%%%%%%%%%%%%%%%%%%%%%%%%%%%%%%%%%%%%%%%%%%%%%%%%%%%%%%%%%%%%%%%%
%%%%%%%%%%%%%%%%%%%%%%%%%%%%%%%%%%%%%%%%%%%%%%%%%%%%%%%%%%%%%%%%%%%%%%%
\section{Open Heavy Flavor}
\label{sec:hf}

\begin{figure}[htb]
	\centering
	\includegraphics[width=0.35\textwidth,valign=c]{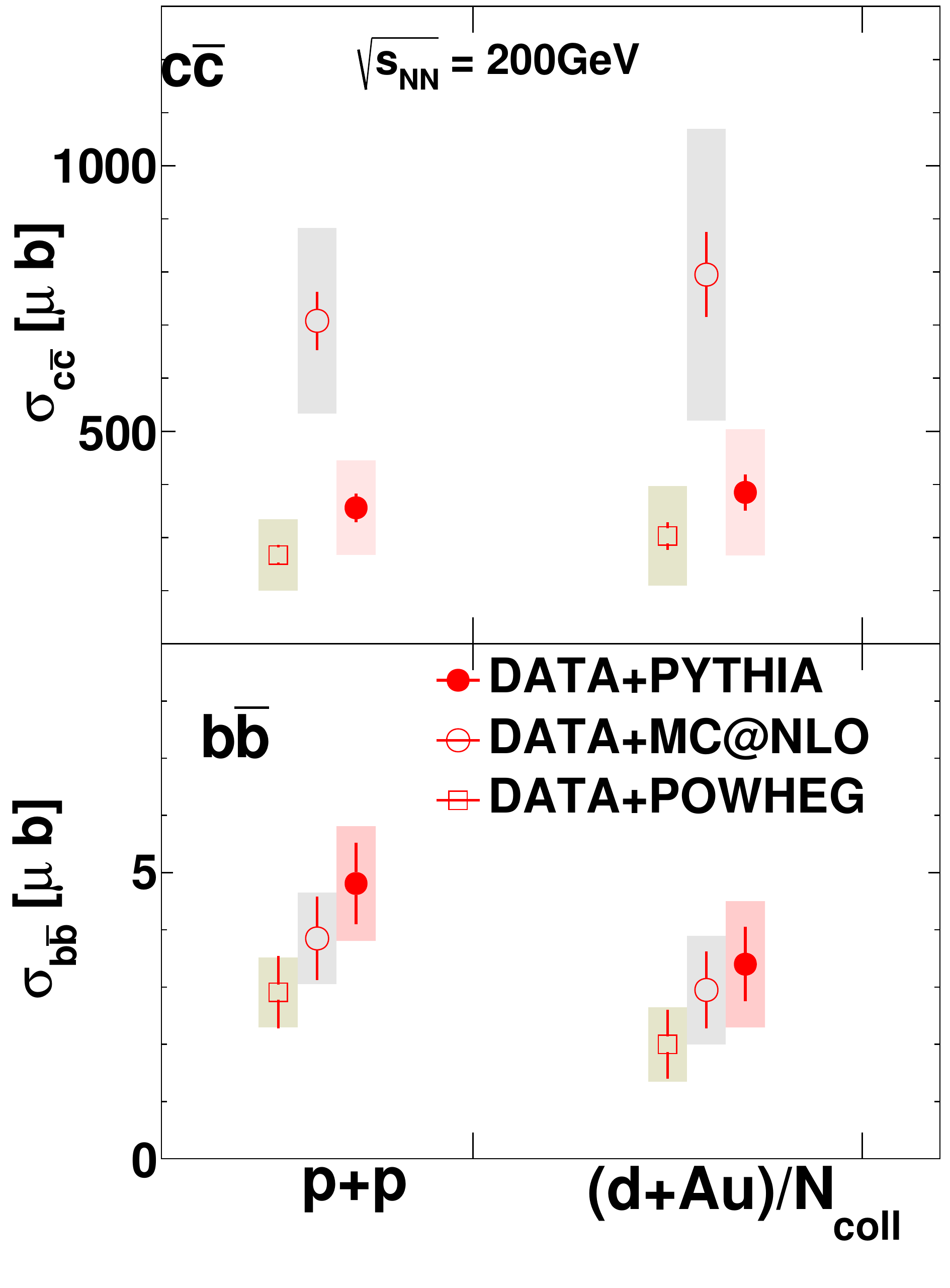}
	\includegraphics[width=0.55\textwidth,valign=c]{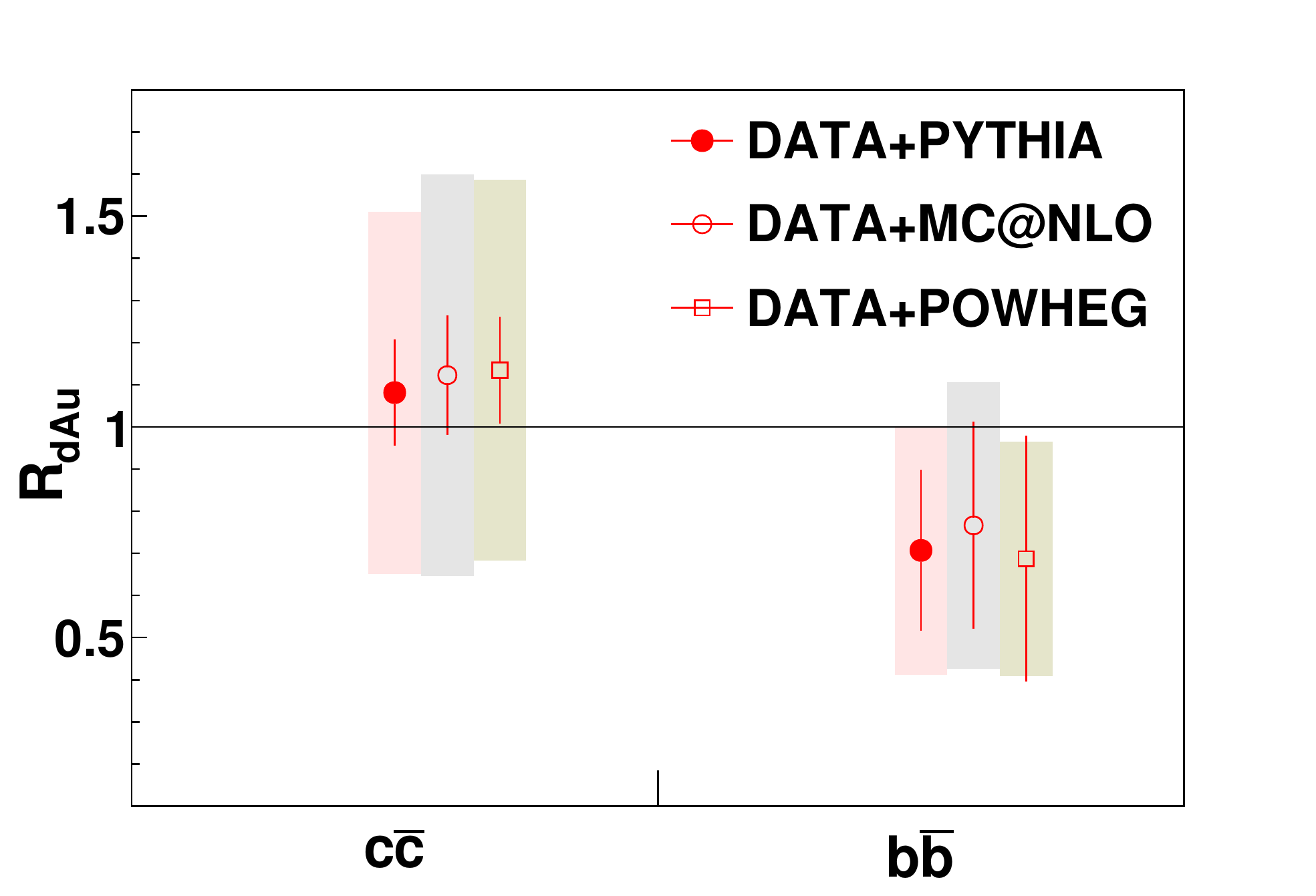}
	\caption{\label{fig:dielectrons} (Left) The $c\bar{c}$ and $b\bar{b}$ cross sections in \pp and \dau collisions at \sqsntwo extrapolated using different methods, as described in the text. (Right) The $R_{dAu}$ for $c\bar{c}$ and $b\bar{b}$ extrapolated using different methods.~\cite{Adare:2017caq}}
\end{figure}

PHENIX has made new measurements of $e^+e^-$ pairs from heavy flavor decays in \pp collisions at \sqsntwo~\cite{Adare:2017caq}. Following similar measurements in \dau collisions~\cite{Adare:2014iwg}, three Monte-Carlo generators are used to simultaniously fit the mass and \pt dependence of the heavy flavor dielectron spectrum in order to separate the $c\bar{c}$ and $b\bar{b}$ components. All three MC generators provide good descriptions of the measured dielectron spectrum. When extrapolating the results to $4\pi$, the results show significant differences between the $c\bar{c}$ and $b\bar{b}$ cross sections from the three generators, as shown in Fig.~\ref{fig:dielectrons}. It is notable that the $c\bar{c}$ cross section shows more variation between the three generators compared to the $b\bar{b}$ cross section, primarily due to uncertainties in the $c\bar{c}$ opening angle. However, when calculating the modification of the $c\bar{c}$ and $b\bar{b}$ cross sections in \dau using the new \pp baseline, the three models give consistent results, which are all consistent with binary scaling of both the charm and bottom.

In recent years PHENIX has made understanding heavy flavor energy loss in the QGP a priority with the installation of two silicon vertex detectors. The midrapidity silicon vertex detector (VTX), installed in 2011, and the forward silicon vertex detector (FVTX), installed in 2012, provide precise vertexing and tracking information.

\begin{figure}[htb]
	\centering
	\includegraphics[width=0.41\textwidth]{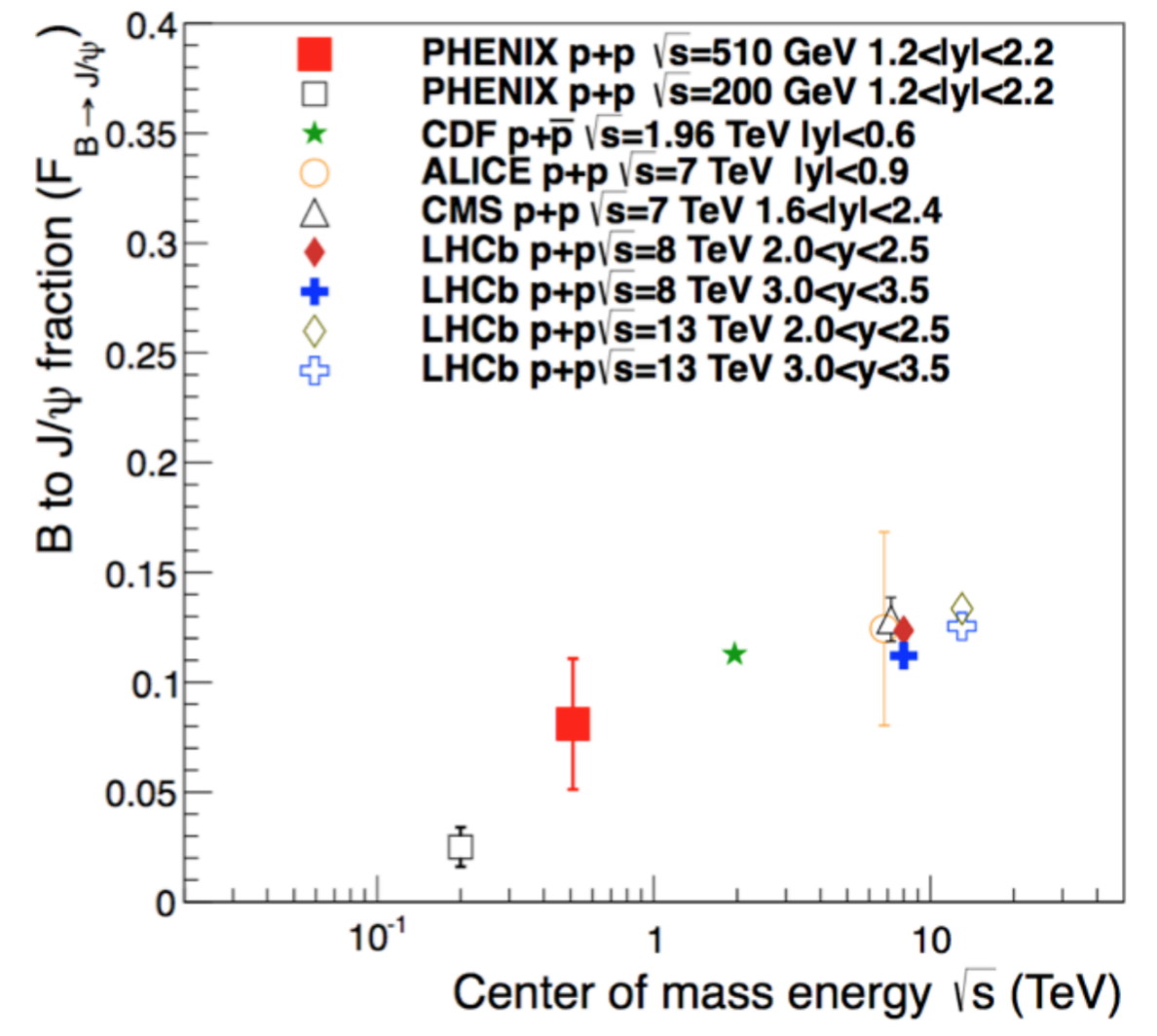}
	\includegraphics[width=0.48\textwidth]{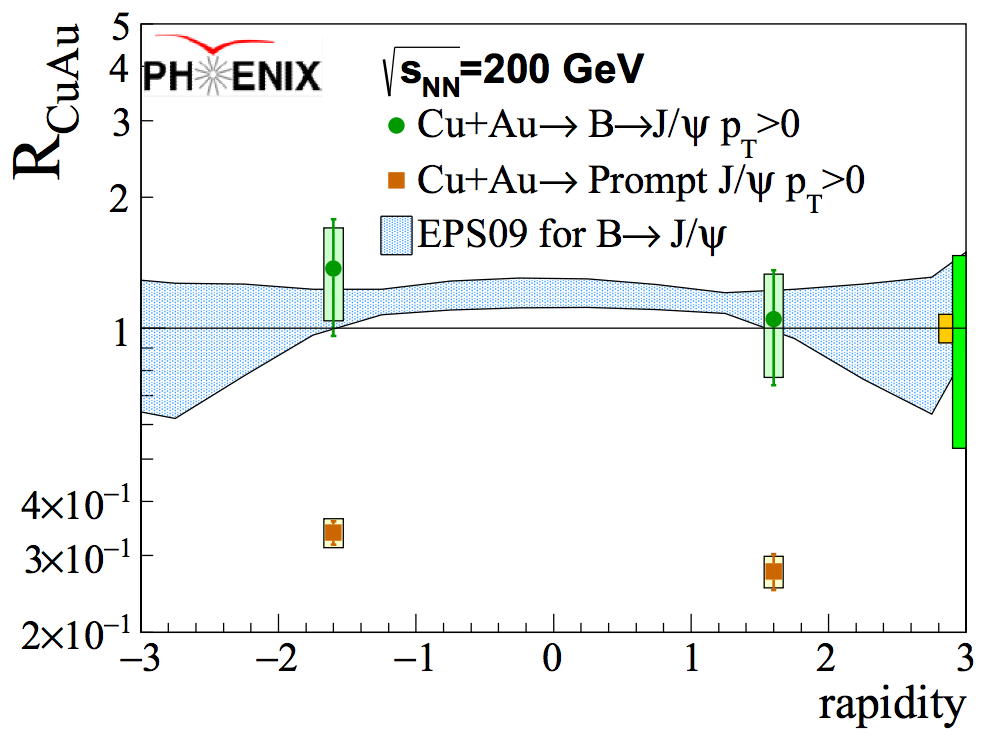}
	\caption{\label{fig:bjpsi} (Left) The fraction of $J/\psi$ from $B$-hadron decays as a function of collision energy. (Right) The nuclear modification factor of $J/\psi$ and $J/\psi$ from $B$ hadron decays in \cuau collisions at \sqsntwo~\cite{Aidala:2017yte}. }
\end{figure}

Using the FVTX, PHENIX has measured the fraction of $J/\psi$'s from $B$-hadron decays, $F_{B\rightarrow J\psi}$, for $J/\psi$'s within $1.2<|y|<2.2$ and $\pt>0$ in \pp collisions at \sqsn~=~510 GeV~\cite{Aidala:2017iad} as well as in \pp and \cuau collisions at \sqsntwo~\cite{Aidala:2017yte}. The results in \pp collisions at \sqsn~=~200 and 510 GeV compared to the world data show a clear energy dependence between 200 GeV and the higher collision energies, as shown in Fig.~\ref{fig:bjpsi}. The nuclear modification factor of $J/\psi$'s from $B$-hadron decays in \cuau, also shown in Fig.~\ref{fig:bjpsi}, is found to be consistent with binary scaling, unlike the prompt $J/\psi$, which are suppressed by more than a factor of three. Within the experimental uncertainties, the $R_{CuAu}$ is also consistent with expectations from nuclear shadowing as given by EPS09~\cite{Eskola:2009uj}. While the statistics of the \cuau data set preclude measurements differential in \pt, the acceptance and efficiency for $B$-hadron decays producing a $J/\psi$ in the PHENIX acceptance is roughly uniform with $B$-hadron \pt and extends down to $\pt=0$.

\begin{figure}[htb]
	\centering
	\includegraphics[width=0.45\textwidth]{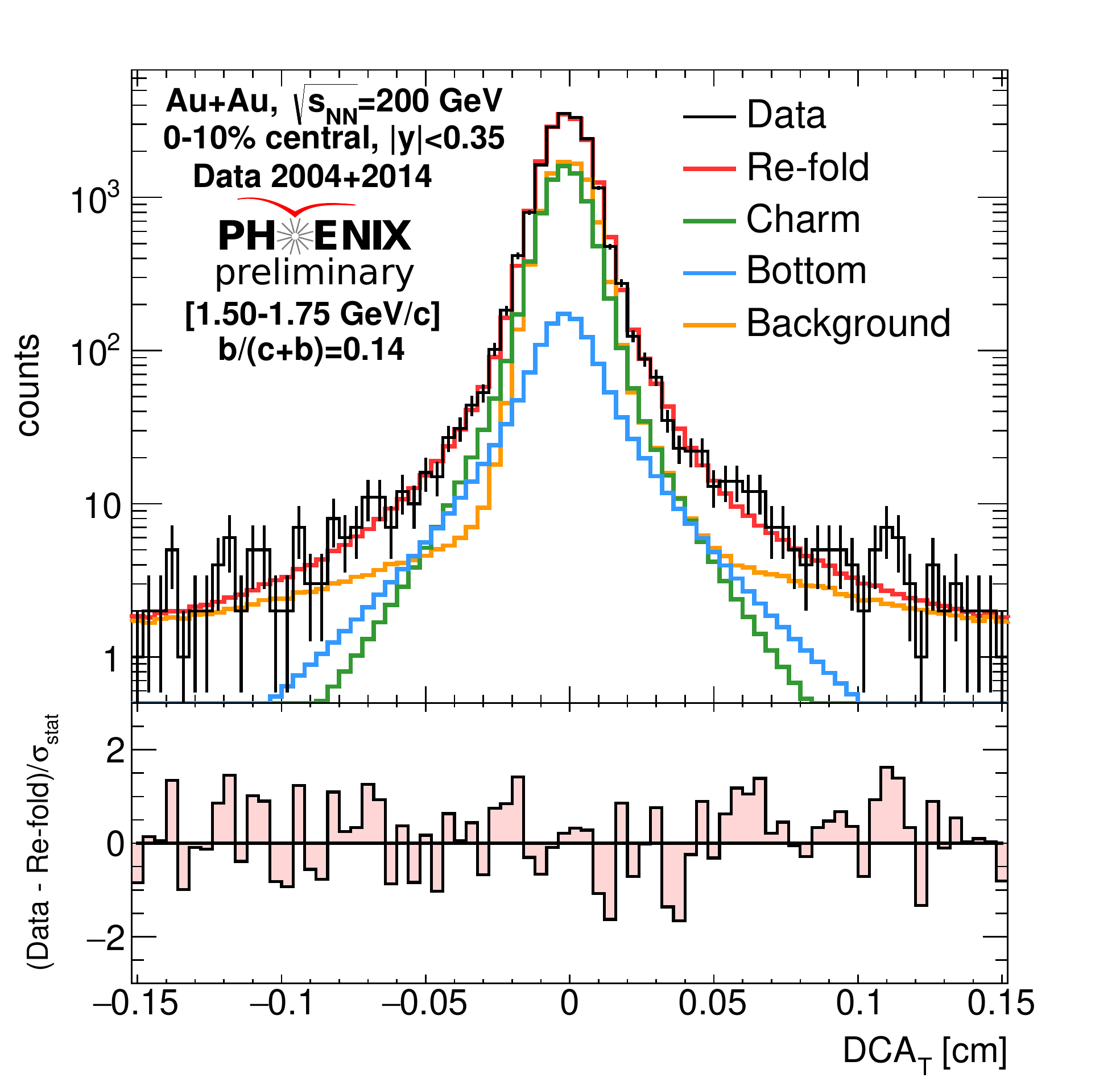}
	\includegraphics[width=0.47\textwidth]{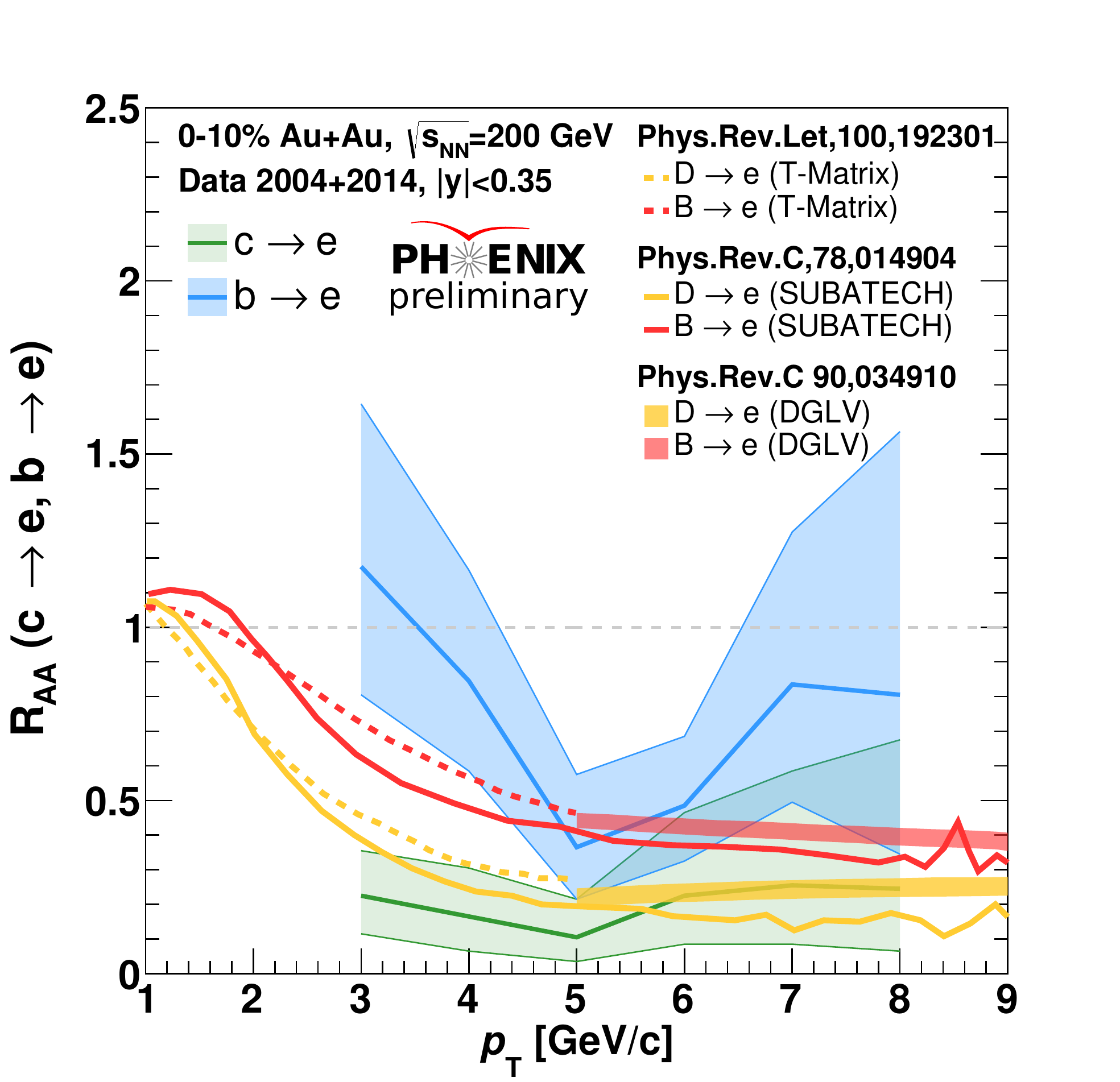}
	\caption{\label{fig:hfe} (Left) The $DCA_T$ distribution for electrons within $1.50<\pt\ [\mathrm{GeV}/c]<1.75$ in 0-10\% central \auau collisions at \sqsntwo. (Right) The $R_{AA}$ for electrons from separated charm and bottom hadron decays in 0-10\% central \auau collisions at \sqsntwo compared to theoretical models discussed in the text.}
\end{figure}

In 2016, PHENIX published it's first measurement of electrons from separated charm and bottom hadron decays in \auau collisions at \sqsntwo using the VTX~\cite{Adare:2015hla} from data collected in 2011. The results indicated that, in MB collisions, electron's from bottom hadron decays are less suppressed than those from charm hadron decays for electron $\pt<4$ GeV/$c$ and similarly suppressed at higher \pt. New preliminary results using the larger statistics available from data recorded in 2014 have measured the nuclear modification factor of these electrons in 0-10\% central \auau collisions at \sqsntwo. The analysis is very similar to that presented in Ref.~\cite{Adare:2015hla}, and uses Bayesian inference to separate electrons from charm and bottom hadron decays using information on the electrons distance of closest approach ($DCA_T$) to the collision vertex. An example $DCA_T$ distribution, along with the corresponding shapes for background electrons and electrons from charm and bottom hadron decays are shown in Fig.~\ref{fig:hfe}. The $R_{AA}$ of electrons from charm and bottom hadron decays, shown in Fig.~\ref{fig:hfe}, indicates a large suppression of electrons from charm which is roughly constant for $\pt>3$ GeV/$c$, and in agreement with calculations from transport~\cite{vanHees:2007me,Gossiaux:2008jv} and energy loss~\cite{Djordjevic:2014yka} models which use large couplings to the medium. Electrons from bottom are seen to be less suppressed for $\pt<5$ GeV/$c$, as expected from transport models. However, the transport models overpredict the suppression of electrons from bottom in this \pt range, hinting that a stronger coupling to the medium might be necessary. For $\pt>5$ GeV/$c$, while the central tendencies of the result indicate that bottom is less suppressed than charm, the current statistical precision of the measurement does not provide an unambiguous ordering of the modifications. The current preliminary result uses only $1/8^\mathrm{th}$ of the available statistics from the combined 2014 and 2016 \auau data samples, and therefore the uncertainties on the modification are expected to be reduced in the final result.

%%%%%%%%%%%%%%%%%%%%%%%%%%%%%%%%%%%%%%%%%%%%%%%%%%%%%%%%%%%%%%%%%%%%%%%
%%%%%%%%%%%%%%%%%%%%%%%%%%%%%%%%%%%%%%%%%%%%%%%%%%%%%%%%%%%%%%%%%%%%%%%
%%%%%%%%%%%%%%%%%%%%%%%%%%%%%%%%%%%%%%%%%%%%%%%%%%%%%%%%%%%%%%%%%%%%%%%
\section{Summary \& Outlook}
\label{sec:summary}

These proceedings have presented a highlighted set of new PHENIX results presented at Quark Matter 2017. New results on flow in small collision systems continue to indicate collective behavior correlated to initial geometry, including at collision energies down to \sqsn~=~19.6 GeV. New measurements of direct photons help constrain the collision energy dependence of the medium temperature. New $\pi^0$ measurements complete a beam species scan with \pdheau collisions and can help improve our understanding of multiple scattering effects. No modification is observed for heavy flavor $e^+e^-$ production in \dau collisions at \sqsntwo. Final results on \pt integrated $B$-hadron modification in \cuau collisions at \sqsntwo, measured via their decay to $J/\psi$ at $1.2<|y|<2.2$, show no modification relative to binary scaling within the experimental uncertainties. Finally, preliminary results on electrons from separated charm and bottom hadron decays in central \auau collisions at \sqsntwo show that electrons from charm are more suppressed than bottom at low-\pt, and that the model calculations including transport and/or energy loss are in reasonable agreement with the data.

After 16 years, PHENIX completed data taking operations at the conclusion of the RHIC 2016 Run. While data taking may have ended, PHENIX looks forward to completing many more analysis. The \sqsntwo \auau data collected in 2014 \& 2016 provide a high statistics sample that should provide definitive measurements of heavy flavor energy loss using the (F)VTX as well as improved direct photon measurements using new methods to tag conversions. The \pdheau and \dau energy scan continue to provide excellent data sets for investigating the limits of collectivity. The 2016 \dau data will also provide a detailed look at gluon shadowing with the newly installed MPC-EX.



%% References


\bibliographystyle{elsarticle-num}
\bibliography{mcglinchey_qm17_proceedings}

\begin{thebibliography}{10}
\expandafter\ifx\csname url\endcsname\relax
  \def\url#1{\texttt{#1}}\fi
\expandafter\ifx\csname urlprefix\endcsname\relax\def\urlprefix{URL }\fi
\expandafter\ifx\csname href\endcsname\relax
  \def\href#1#2{#2} \def\path#1{#1}\fi

\bibitem{CMS:2012qk}
S.~Chatrchyan, et~al., {Observation of long-range near-side angular
  correlations in proton-lead collisions at the LHC}, Phys.Lett. B718 (2013)
  795--814.
\newblock \href {http://dx.doi.org/10.1016/j.physletb.2012.11.025}
  {\path{doi:10.1016/j.physletb.2012.11.025}}.

\bibitem{Abelev:2012ola}
B.~Abelev, et~al., {Long-range angular correlations on the near and away side
  in $p$-Pb collisions at $\sqrt{s_{NN}}=5.02$ TeV}, Phys.Lett. B719 (2013)
  29--41.
\newblock \href {http://dx.doi.org/10.1016/j.physletb.2013.01.012}
  {\path{doi:10.1016/j.physletb.2013.01.012}}.

\bibitem{Aad:2012gla}
G.~Aad, et~al., {Observation of Associated Near-Side and Away-Side Long-Range
  Correlations in $\sqrt{s_{NN}}$=5.02  TeV Proton-Lead Collisions with the
  ATLAS Detector}, Phys. Rev. Lett. 110~(18) (2013) 182302.
\newblock \href {http://dx.doi.org/10.1103/PhysRevLett.110.182302}
  {\path{doi:10.1103/PhysRevLett.110.182302}}.

\bibitem{Adare:2013piz}
A.~Adare, et~al., {Quadrupole Anisotropy in Dihadron Azimuthal Correlations in
  Central $d$$+$Au Collisions at $\sqrt{s_{_{NN}}}$=200 GeV}, Phys. Rev. Lett.
  111~(21) (2013) 212301.
\newblock \href {http://dx.doi.org/10.1103/PhysRevLett.111.212301}
  {\path{doi:10.1103/PhysRevLett.111.212301}}.

\bibitem{Adare:2014keg}
A.~Adare, et~al., {Measurement of long-range angular correlation and quadrupole
  anisotropy of pions and (anti)protons in central $d$$+$Au collisions at
  $\sqrt{s_{_{NN}}}$=200 GeV}, Phys. Rev. Lett. 114~(19) (2015) 192301.
\newblock \href {http://dx.doi.org/10.1103/PhysRevLett.114.192301}
  {\path{doi:10.1103/PhysRevLett.114.192301}}.

\bibitem{Adare:2015ctn}
A.~Adare, et~al., {Measurements of elliptic and triangular flow in
  high-multiplicity $^{3}$He$+$Au collisions at $\sqrt{s_{_{NN}}}=200$ GeV},
  Phys. Rev. Lett. 115~(14) (2015) 142301.
\newblock \href {http://dx.doi.org/10.1103/PhysRevLett.115.142301}
  {\path{doi:10.1103/PhysRevLett.115.142301}}.

\bibitem{Aidala:2016vgl}
C.~Aidala, et~al., {Measurement of long-range angular correlations and
  azimuthal anisotropies in high-multiplicity $p$$+$Au collisions at
  $\sqrt{s_{_{NN}}}=200$ GeV}, Phys. Rev. C95~(3) (2017) 034910.
\newblock \href {http://dx.doi.org/10.1103/PhysRevC.95.034910}
  {\path{doi:10.1103/PhysRevC.95.034910}}.

\bibitem{Nagle:2013lja}
J.~L. Nagle, A.~Adare, S.~Beckman, T.~Koblesky, J.~Orjuela~Koop, D.~McGlinchey,
  P.~Romatschke, J.~Carlson, J.~E. Lynn, M.~McCumber, {Exploiting Intrinsic
  Triangular Geometry in Relativistic He3+Au Collisions to Disentangle Medium
  Properties}, Phys. Rev. Lett. 113~(11) (2014) 112301.
\newblock \href {http://dx.doi.org/10.1103/PhysRevLett.113.112301}
  {\path{doi:10.1103/PhysRevLett.113.112301}}.

\bibitem{Habich:2014jna}
M.~Habich, J.~L. Nagle, P.~Romatschke, {Particle spectra and HBT radii for
  simulated central nuclear collisions of C + C, Al + Al, Cu + Cu, Au + Au, and
  Pb + Pb from $\sqrt{s}=62.4$ - $2760$ GeV}, Eur. Phys. J. C75~(1) (2015) 15.
\newblock \href {http://dx.doi.org/10.1140/epjc/s10052-014-3206-7}
  {\path{doi:10.1140/epjc/s10052-014-3206-7}}.

\bibitem{Koop:2015trj}
J.~D. Orjuela~Koop, R.~Belmont, P.~Yin, J.~L. Nagle, {Exploring the Beam Energy
  Dependence of Flow-Like Signatures in Small System $d+$Au Collisions}, Phys.
  Rev. C93~(4) (2016) 044910.
\newblock \href {http://dx.doi.org/10.1103/PhysRevC.93.044910}
  {\path{doi:10.1103/PhysRevC.93.044910}}.

\bibitem{Romatschke:2015gxa}
P.~Romatschke, {Light-Heavy Ion Collisions: A window into pre-equilibrium QCD
  dynamics?}, Eur. Phys. J. C75~(7) (2015) 305.
\newblock \href {http://dx.doi.org/10.1140/epjc/s10052-015-3509-3}
  {\path{doi:10.1140/epjc/s10052-015-3509-3}}.

\bibitem{Lin:2004en}
Z.-W. Lin, C.~M. Ko, B.-A. Li, B.~Zhang, S.~Pal, {A Multi-phase transport model
  for relativistic heavy ion collisions}, Phys. Rev. C72 (2005) 064901.
\newblock \href {http://dx.doi.org/10.1103/PhysRevC.72.064901}
  {\path{doi:10.1103/PhysRevC.72.064901}}.

\bibitem{Ollitrault:2009ie}
J.-Y. Ollitrault, A.~M. Poskanzer, S.~A. Voloshin, {Effect of flow fluctuations
  and nonflow on elliptic flow methods}, Phys. Rev. C80 (2009) 014904.
\newblock \href {http://dx.doi.org/10.1103/PhysRevC.80.014904}
  {\path{doi:10.1103/PhysRevC.80.014904}}.

\bibitem{Adare:2015gla}
A.~Adare, et~al., {Centrality-dependent modification of jet-production rates in
  deuteron-gold collisions at $\sqrt{s_{NN}}$=200 GeV}, Phys. Rev. Lett.
  116~(12) (2016) 122301.
\newblock \href {http://dx.doi.org/10.1103/PhysRevLett.116.122301}
  {\path{doi:10.1103/PhysRevLett.116.122301}}.

\bibitem{McGlinchey:2016ssj}
D.~McGlinchey, J.~L. Nagle, D.~V. Perepelitsa, {Consequences of high-$x$ proton
  size fluctuations in small collision systems at $\sqrt {s_{NN}}=$ 200GeV},
  Phys. Rev. C94~(2) (2016) 024915.
\newblock \href {http://dx.doi.org/10.1103/PhysRevC.94.024915}
  {\path{doi:10.1103/PhysRevC.94.024915}}.

\bibitem{Adare:2017caq}
A.~Adare, et~al., {Measurements of $e^+e^-$ pairs from open heavy flavor in
  $p$+$p$ and $d$+$A$ collisions at $\sqrt{s_{NN}}=200$ GeV}\href
  {http://arxiv.org/abs/1702.01084} {\path{arXiv:1702.01084}}.

\bibitem{Adare:2014iwg}
A.~Adare, et~al., {Cross section for $b\bar{b}$ production via dielectrons in
  d$+$Au collisions at $\sqrt{s_{_{NN}}}=200$ GeV}, Phys. Rev. C91~(1) (2015)
  014907.
\newblock \href {http://dx.doi.org/10.1103/PhysRevC.91.014907}
  {\path{doi:10.1103/PhysRevC.91.014907}}.

\bibitem{Aidala:2017iad}
C.~Aidala, et~al., {Measurements of $B \rightarrow J/\psi$ at forward rapidity
  in $p$$+$$p$ collisions at $\sqrt{s}=510$ GeV}\href
  {http://arxiv.org/abs/1701.01342} {\path{arXiv:1701.01342}}.

\bibitem{Aidala:2017yte}
C.~Aidala, et~al., {B-meson production at forward and backward rapidity in
  $p$+$p$ and Cu+Au collisions at $\sqrt{s_{_{NN}}}$=200 GeV}\href
  {http://arxiv.org/abs/1702.01085} {\path{arXiv:1702.01085}}.

\bibitem{Eskola:2009uj}
K.~J. Eskola, H.~Paukkunen, C.~A. Salgado, {EPS09: A New Generation of NLO and
  LO Nuclear Parton Distribution Functions}, JHEP 04 (2009) 065.
\newblock \href {http://dx.doi.org/10.1088/1126-6708/2009/04/065}
  {\path{doi:10.1088/1126-6708/2009/04/065}}.

\bibitem{Adare:2015hla}
A.~Adare, et~al., {Single electron yields from semileptonic charm and bottom
  hadron decays in Au$+$Au collisions at $\sqrt{s_{NN}}=200$ GeV}, Phys. Rev.
  C93~(3) (2016) 034904.
\newblock \href {http://dx.doi.org/10.1103/PhysRevC.93.034904}
  {\path{doi:10.1103/PhysRevC.93.034904}}.

\bibitem{vanHees:2007me}
H.~van Hees, M.~Mannarelli, V.~Greco, R.~Rapp, {Nonperturbative heavy-quark
  diffusion in the quark-gluon plasma}, Phys.Rev.Lett. 100 (2008) 192301.
\newblock \href {http://dx.doi.org/10.1103/PhysRevLett.100.192301}
  {\path{doi:10.1103/PhysRevLett.100.192301}}.

\bibitem{Gossiaux:2008jv}
P.~B. Gossiaux, J.~Aichelin, {Towards an understanding of the RHIC single
  electron data}, Phys. Rev. C78 (2008) 014904.
\newblock \href {http://arxiv.org/abs/0802.2525} {\path{arXiv:0802.2525}},
  \href {http://dx.doi.org/10.1103/PhysRevC.78.014904}
  {\path{doi:10.1103/PhysRevC.78.014904}}.

\bibitem{Djordjevic:2014yka}
M.~Djordjevic, M.~Djordjevic, {Heavy flavor puzzle from data measured at the
  BNL Relativistic Heavy Ion Collider: Analysis of the underlying effects},
  Phys.Rev. C90~(3) (2014) 034910.
\newblock \href {http://dx.doi.org/10.1103/PhysRevC.90.034910}
  {\path{doi:10.1103/PhysRevC.90.034910}}.

\end{thebibliography}






% \end{thebibliography}

\end{document}